\documentclass[preprint,12pt]{elsarticle}

\usepackage{amssymb}
\usepackage{url}
\usepackage{bm}
\usepackage[ruled, vlined,linesnumbered]{algorithm2e}
\usepackage{algorithmic}
\usepackage[bookmarks]{hyperref}
\usepackage{float}
\usepackage{verbatim}

\def\mbf#1{\mathbf{#1}}

\journal{}

\begin{document}

\begin{frontmatter}




\title{Accelerating DEM simulations on GPUs by reducing the impact of warp divergences}


\author[rvt]{Y.~Nakahara}
\ead{nakahara.yasuhiro105@canon.co.jp}

\author[rvt]{T.~Washizawa}
\ead{washizawa.teruyoshi@canon.co.jp}

\address[rvt]{Simulation \& Analysis R\&D Center, Canon Inc., 3-30-2, Shimomaruko, Ohta-ku, Tokyo 146-8501, Japan}

\begin{abstract}
A way to accelerate DEM calculations on the GPUs is developed. We examined how warp divergences take place in the contact detection and the force calculations taking account of the GPU architecture. Then we showed a strategy to reduce the impact of the warp divergences on the runtime of the DEM force calculations.
\end{abstract}


\end{frontmatter}


\section{Introduction}
\label{}
Many kinds of applications for scientific simulations such as matrix computations or computations of continuous models on uniform grids have been successfully applied to GPU (Graphics Processing Unit)~\cite{GPU_Applications}. The speed-up by the GPU depends on applications.

In this paper, we focus on the distinct element method or DEM for short~\cite{cundall1979discrete}. This method provides simulations of powder or many tiny particles of a solid substance.
The simulations based on DEM are applied to the engineering involving the material engineering, medicine manufacturing, civil engineering and food engineering.
Here we discuss a DEM program mainly applied to motion of toners used in the electric photo processing.

In order to apply GPUs to DEM simulations, some optimization techniques are required.
Particle simulations based on DEM are likely to access memory in a random way. They also have conditional branches, which cause warp divergences.
The random accesses and the conditional branches will limit the effective performance of the GPU.

In this work we focus on the reduction of the performance limitation of DEM force calculations caused by the warp divergences.

\section{Particle simulations based on DEM}
\label{}
In the DEM, the computation of the contact force acting on a particle is rather complicated. The force is not defined by a simple potential function. The computation is carried out by using the following equations:
\begin{equation}
\mathbf{F} = -k_t \boldsymbol\delta_t - \eta_t \mathbf{v}_t - k_n \delta_n^{3/2} \mathbf{n} -\eta_n \mathbf{v}_n , \label{equation-calcF} 
\end{equation}
\begin{equation}
\mathbf{T} = r \left(\mathbf{n} \times \mathbf{F} \right), \nonumber
\end{equation}
where $\mathbf{F}$ is the forces acting between two particles, $\mathbf{T}$ is the torque caused by the force $\mathbf{F}$. $\mathbf{v}$ is the relative velocity between two particles at their contact point. $r$ is the radius of a particle. $\mathbf{n}$ is the unit vector from the center of a particle to that of the other particle. $k$ and $\eta$ are the spring coefficient and damping coefficient.
$\boldsymbol\delta$ shows the displacement between two particles measured after they become in contact. $\boldsymbol\delta$ means how a particle is deformed by another particle. The definition of the displacement is shown below.
The suffixes of $n$ and $t$ show the component of vectors in the normal direction and tangential direction respectively.

The tangential components of $\mathbf{v}$ and $\boldsymbol\delta$ are calculated, respectively, by the following equations:
\begin{equation}
\mathbf{v}_t = \mathbf{v} - (\mathbf{v} \cdot \mathbf{n}) \mathbf{n} + ( r_1 \boldsymbol\omega_1 + r_2 \boldsymbol\omega_2 ) \times \mathbf{n} , \label{equation-vt} 
\end{equation}
\begin{equation}
\boldsymbol\delta_t = \boldsymbol\delta_{t, old} - \left(  \boldsymbol\delta_{t, old} \cdot \mbf{n} \right) \mbf{n} + \mbf{v}_t \Delta t. \label{equation-dt}
\end{equation}
We use the suffixes of $1$ and $2$ to indicate each of two particles.
The coefficients, $k$ and $\eta$, in Eq.~(\ref{equation-calcF}) are defined by the following equations:

\begin{equation}
k_t = 8 \sqrt{\frac{r_1 r_2}{r_1 + r_2} \delta_n} \left( \frac{2-\sigma_1}{G_1} + \frac{2-\sigma_2}{G_2} \right)^{-1} , \nonumber
\end{equation}
\begin{equation}
k_n = \frac{4}{3} \sqrt{\frac{r_1 r_2}{r_1 + r_2}} \left( \frac{2-\sigma_1^2}{E_1} + \frac{2-\sigma_2^2}{E_2} \right)^{-1} , \nonumber
\end{equation}
\begin{equation}
\eta_n = \eta_t = \alpha(\epsilon_{1,2}) \sqrt{\frac{m_1 m_2}{m_1+m_2}k_n \sqrt{\delta_n}} , \nonumber
\end{equation}
where $\sigma$, $G$, $E$ and $m$ are the Poisson ratio, transverse elastic modulus, young ratio and mass, respectively.
$\alpha$ is the restitution parameter, which is a function of the coefficient of restitution, $\epsilon_{1,2}$.

Taking account of the angular velocities of the two particles, Eq.~(\ref{equation-vt}) calculates the tangential component of the relative velocity between two particles at their contact point.
Eq.~(\ref{equation-dt}) calculates the displacement between two particles in the tangential direction, $\boldsymbol\delta_t$, by integrating the relative velocity of two particles over the period during which they are in contact.
The displacement in the normal direction, $\boldsymbol\delta_n$, is defined to be the overlap of the particles in the normal direction and then simply calculated from the particles' center positions and their radii.

In addition, in order to take account of sliding friction, if the magnitude of calculated DEM forces acting between two particles in the tangential direction is greater than the threshold, $\mu_D |\mbf{F}_n|$, the DEM force in the tangential direction is replaced with the sliding friction defined by the following equation:
\begin{equation}
\mbf{F}_{t, new} = \mu_D |\mbf{F}_n| \left( \mbf{F}_t / |\mbf{F}_t| \right) , \label{equation-slidingFriction}
\end{equation}
where $\mu_D$ is the coefficient of sliding friction. After the forces in the tangential direction are replaced, $\boldsymbol\delta_t$ is recalculated so that it satisfies $\mbf{F}_{t, new} = - k_t \boldsymbol\delta_t$.

A DEM based program needs variables depending on two particles to store the displacement in the tangential direction.
These variables are stored in the condensed format of a sparse matrix in our DEM code.
It will take some computational cost to search the value associated with two particle IDs from the memory.

In order to reduce the amount of the computation, it is important to effectively find particles which contact with a particle of interest. One of the methods to find neighbor particles effectively is the cell-based method. This method creates uniform cells and assigns each particle to the associated cell. We can find candidates for contact particles effectively by searching the neighbor cells of the cell containing a particle of interest.

\begin{figure}[H]
\centering
\includegraphics[width=120mm]{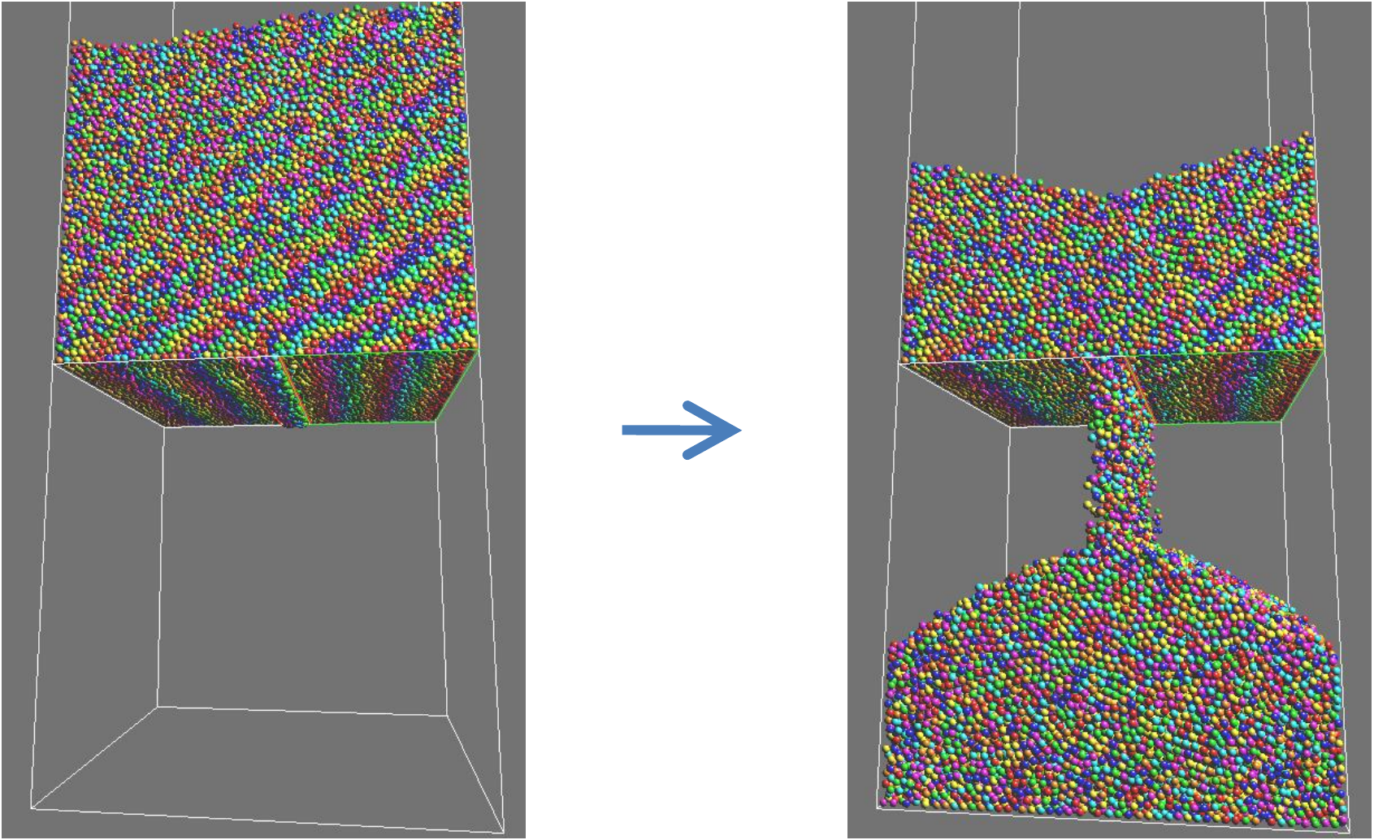}
\caption{Some views of DEM calculation results}\label{figure-simulation}
\end{figure}

\section{Optimization for GPUs}
\label{}

\subsection{Previous work}
\label{}

The NVIDIA's sample code, "oclParticles"~\cite{GPU_Computing_SDK}, shows the way to manage the data of particles effectively in a GPU. oclParticles makes the search of neighbor particles faster by using the cell-based method mentioned in the previous section. In addition, in order to accelerate the access to the memory, the code sorts the particle data in the cell order at each time step. NVIDIA provides a fast sorting method based on the parallel bitonic sort.

A simplified DEM, where particles do not have the attribute of the angular velocity, was implemented on a GPU~\cite{Harada2007}. The NVIDIA's "oclParticles" is also based on a simplified DEM. A DEM based two-dimensional powder calculation was implemented on a GPU~\cite{Shigeto2008758}. An algorithm making the DEM based three-dimensional simulations faster was applied to multi-core processors including GPUs~\cite{Shigeto2011398}. An algorithm to calculate the DEM forces with the action-reaction law being applied was developed~\cite{Nishiura20111923}.

Particle based simulations such as SPH (Smoothed Particle Hydrodynamics) or PIC (Particle-in-Cell) also have problems similar to DEM. Like DEM particle simulations, SPH programs calculate the particle interaction between two particles only if the distance between two particles is less than a given value. Some optimization techniques for searching contact pairs in SPH on a GPU have been developed~\cite{Dominguez2013617}. In applying a GPU to PIC programs, which are mainly used in plasma simulations, it was reported that keeping particles ordered in the memory improves the performance~\cite{Decyk2011641}.

\subsection{Characteristics of the GPU}
\label{}
A GPU processor has many primitive arithmetic operation units. NVIDIA calls the processor SM (Streaming Multiprocessor). The primitive arithmetic operation unit is called CUDA (Compute Unified Device Architecture) core. GPU programmers can write their programs in the way that these CUDA cores work independently by using the CUDA programming environment.

NVIDIA calls a group of 32 successive GPU threads a warp. Each GPU's thread in a warp is not independent with respect to the activity of the associated hardware. The threads in a warp cannot execute different instructions at the same time. If all the threads in a warp execute one instruction stored at the same address, the operations will be carried out at the same time. Since each thread has different local stores such as registers and uses them as the operand of an instruction,  each thread in a warp modifies different registers or global memory at the execution of an instruction. This makes the GPU calculation faster. This means that GPU threads in a warp execute arithmetic operations in parallel in a SIMD (Single Instruction Multiple Data) manner. This kind of architecture is called SIMT (Single Instruction Multi Threads).

If a program has many branches and the threads executing the program take different branches, not all the threads in a warp will require the same instruction. Since a warp serially executes instructions, the threads in a warp can not execute instructions stored at different addresses at the same time. In this case, the GPU has an overhead. This is called the warp divergence.

In addition, it should be noticed that calculation time depends on how each of the threads in a warp accesses the memory, even though all the threads in a warp execute the same instruction. For example, if successive threads in a warp access continuously stored data, the execution time will be smaller.

If-statements in the GPU source code cause branches in the GPU instructions. Branches will also take place implicitly in the statements such as for-loops.
The impact of the overhead of the warp divergences on the total calculation time depends on the algorithm.

\subsection{Implementation of DEM on the GPU before optimization}
In order to implement the DEM on the GPU, we started with the NVIDIA's sample code, "oclParticles".
Since its equations for force calculations are too simple, we added the required functions calculating the forces acting on particles according to Eq.~(\ref{equation-calcF}) before the optimization that we are going to suggest in this paper.
The implementations before the optimization are as follows.

\begin{enumerate}[1.]
\item We added the variables required in this calculations. Here, we added the variable of the displacement between two particles in the tangential direction, $\boldsymbol\delta_t$, which is a specific variable in the DEM.

\item We added the functions reordering the memory of added variables in the cell order.

\item We added the function calculating the forces acting on the particles. We implemented this according to Eq.~(\ref{equation-calcF}) and Eq.~(\ref{equation-slidingFriction}). The torque acting on the particles is also calculated.

\item In the function updating the attributes of particles, we modified the code so that the angular velocities of the particles are calculated.

\item We added variables of walls. The forces acting between a particle and a wall are calculated in the same way as the forces acting between particles. We implemented rectangles and straight lines of walls.
\end{enumerate}

In this code, each thread is associated with a particle. Each thread finds the neighbor particles and calculates the DEM forces if two particles are in contact. This is carried out in the following way.

\begin{enumerate}[Step 1.]
\item A thread finds the cell containing a particle of interest.
\item The thread gets the neighbor cells.
\item Among particles in the cell the thread tries finding particles in contact with the thread-associated particle
\item If two particles in contact are found, the thread calculates the forces acting between two particles.
\end{enumerate}

The detailed flow of the DEM force calculations is shown in Algorithm \ref{algorithm-dem-force}.

\begin{algorithm}[H]
\DontPrintSemicolon 
$i \leftarrow $ the thread index \;
$p1 \leftarrow $ the thread-associated particle\;
$c \leftarrow $ the particle-associated cell \;
\ForEach{neighbor cell $nc \in$ the neighbor cells of $c$} {
  \ForEach{particle $p2 \in$ the particles in $nc$} {
    \If{$p1$ and $p2$ are in contact} {
      Update $\boldsymbol\delta_{t}(p1, p2) \leftarrow \boldsymbol\delta_{t,old}(p1, p2)$ \;
      Calculate DEM forces $\mbf{F}(p1, p2)$ \;
      \If{$|\mbf{F}_t(p1, p2)| > \mu_D |\mbf{F}_n(p1, p2)|$} {
        $\mbf{F}_t(p1, p2)\!\leftarrow \mu_D |\mbf{F}_n(p1, p2)| \frac{ \mbf{F}_t(p1, p2) }{ |\mbf{F}_t(p1, p2)| }$ \;
      }
      Calculate torque $\mbf{T}(p1, p2)$ \;
      Update forces : $\mbf{F}(p1)  \leftarrow \mbf{F}(p1) + \mbf{F}(p1, p2)$ \;
      Update torque : $\mbf{T}(p1)  \leftarrow \mbf{T}(p1) + \mbf{T}(p1, p2)$ \;
    }
  }
}
\caption{Calculations of the DEM forces acting on particles}\label{algorithm-dem-force}
\end{algorithm}

\vspace{1em}

The particles are stored in the global memory in the cell order, and successively indexed particles are often in the same cell. Since the number of threads in a warp is 32 or more, and the number of the particles in a cell will be less than 32, a warp will be associated with at least two cells containing different numbers of particles. Therefore the for-loop in the code finding contact candidates will involve warp divergences. Though smaller cell size will make finding contact particles faster, it will cause more divergences. Furthermore, too small cell size will need next-neighbor search.

In addition, each thread checks if the thread-associated particle and each of the contact candidates are in contact. If the two particles are actually in contact, the thread calculates the force acting on the thread-associated particle. This part of the code also causes warp divergences.

Note that we do not make use of the symmetry of the force or the action-reaction law. This avoids costly exclusive modification of the global memory.

\subsection{Optimization strategy for DEM force calculations}
\label{}
\begin{figure}[H]
\centering
\includegraphics[width=135mm]{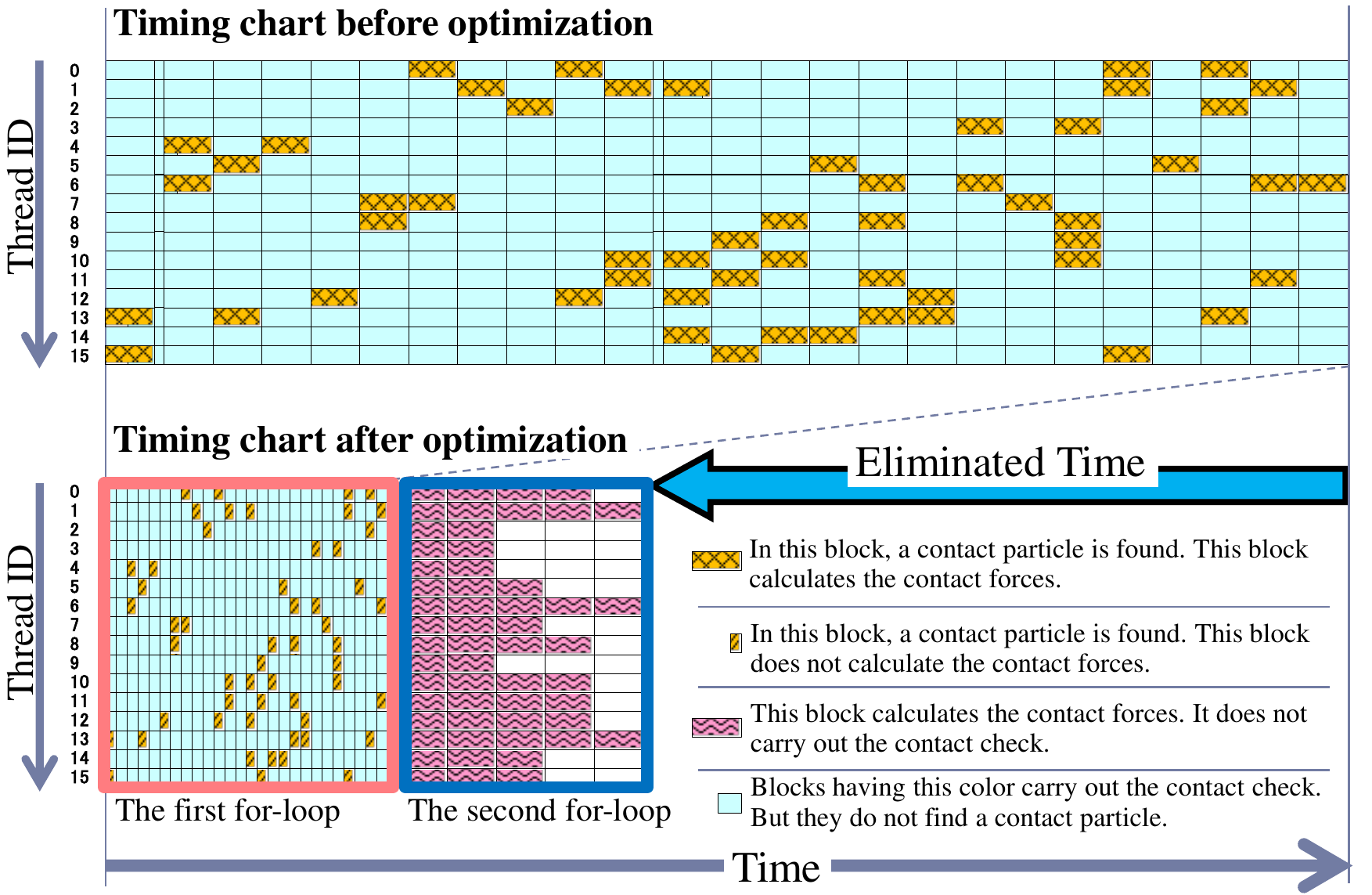}
\caption{Simply illustrated timing charts used for explaining how the threads in a warp work in the execution of the DEM force calculations.}\label{figure-time-chart}
\end{figure}

We show the impact of the warp divergences on the total DEM force calculation time and suggest the way to reduce the overhead of the warp divergences by using Fig.~\ref{figure-time-chart}.

In this figure, we are thinking of how threads in a warp work in the DEM force calculations on the GPU.

In Fig.~\ref{figure-time-chart} the vertical axis and horizontal axis indicate the thread ID in a warp and elapsed time respectively. A rectangle-shaped block is associated with the loop block corresponding to the lines 5--13 in Algorithm \ref{algorithm-dem-force}. Since the thread accesses $27(=3^3)$ cells to find neighbor particles, it carries out more than 27 loop blocks.  The width of a rectangle-shaped block means the elapsed time required to carry out the block. In this figure, for simplicity, we assume that a cell includes at most one particle. In addition, for the thread ID in the vertical axis, we display only 16 threads. The actual number of threads in a warp is 32 in the GPU we used.

The upper timing chart of Fig.~\ref{figure-time-chart} is the one before optimization.
In each loop block of this chart, the thread checks if the thread-associated particle and each of the particles in the cell are in contact. The thread calculates the DEM force if the two particles are actually in contact. If a thread carries out the DEM force calculation the associated block will cost larger elapsed time.
An orange-colored block filled with trellises shows a loop block in which two particles are in contact.
This chart illustrates that the majority of the threads are inactive and waiting for the minority of the threads to finish the DEM force calculations. This means that the associated CUDA cores are inefficiently used.

Now, we suggest dividing the for-loop statement for the DEM force calculations into two for-loops. We do not calculate the DEM force soon after finding a contact particle. Instead, we delay the force calculations. Before the force calculations, each thread finds all particles in contact with the thread-associated particle. Each thread keeps the contact particle IDs in the first for-loop.
In the second for-loop, the program concentrates on the DEM force calculations by using the obtained particle IDs.

After applying this optimization, we will get the lower timing chart of Fig.~\ref{figure-time-chart}. The loop blocks in the first-loop only check the contacts. The elapsed time to check the contact is much smaller than that of the force calculation. Therefore, the elapsed time of each thread over the first-loop becomes much smaller.

The CUDA cores must be more efficiently used in the second loop. The number of the loop blocks in the second for-loop must be around six, which is less than the number of the loop blocks in the first-loop. The lower timing chart of Fig.~\ref{figure-time-chart} illustrates that the proportion of red-colored blocks filled with waves to all the blocks of the second-loop becomes large.

Here, we assumed that the radii of the particles are the same or nearly the same. In this case, the number of the particles in contact with each particle is at most six or so. Therefore, the number of the loop blocks in the second for-loop must be up to around six. If the maximum number of contact particles in a warp is greater than the number of the loop blocks in the first for-loop, the DEM force calculation time of the optimized code will become comparable to or greater than that of the pre-optimized code. Thus, the number of contact particles affects the effectiveness of our method.

\section{Details of implementation}
\label{}
\subsection{Programming environments}
\label{}
NVIDIA provides some programming environments for the GPUs. In the environments, the CUDA language is mainly used. The language specification of CUDA is based on the C language. GPU code is combined with CPU code. The CUDA compiler can make an executable binary file for the GPU and the CPU from the same source code. So, CUDA will facilitate the GPU programming.

In this work, however, we took OpenCL (Open Computing Language)~\cite{OpenCL_Specification} taking account of the portability. In the OpenCL programming environment, GPU code is not combined with CPU code. The OpenCL environment is provided in the form of libraries. Executable CPU binary code linked with the OpenCL libraries reads the OpenCL source code written in different files and creates executable GPU binary code.

The source code written in CUDA is compiled at the same time when the CPU source code is compiled. So, the CUDA GPU compiler and the CPU compiler such as gcc are tightly bound. In case of OpenCL, we can take any CPU compiler as long as the code can link the OpenCL library. This is an advantage of OpenCL. A disadvantage is that OpenCL may not be able to  access all the advanced architectures of the NVIDIA GPUs.

\subsection{Implementation before optimization}
\label{}
In order to search the neighbor particles fast, we made use of an algorithm used in the NVIDIA's sample code, “oclParticles”. This sample code has a sorting program. This is based on the algorithm of the bitonic sort and well optimized for the GPU.

The equation used to calculate the forces acting between particles in the NVIDIA's code is different from ours. The NVIDIA's code uses a simple equation to calculate the forces. We took more realistic equations to obtain precise result. We modified the source code calculating the forces acting on particles according to Eq.~(\ref{equation-calcF}).
We added the required attributes to the particles. One of the added attributes is the displacement between two particles in the tangential direction. This is associated with the contact pair. We stored the data in the global memory in the manner that a sparse matrix is stored.

All the kernels used in this calculation are shown below. The only process carried out on the CPU is creating initial configurations of particles. After the CPU sends the initial configurations to the GPU, all the calculations are carried out on the GPU. The CPU only iteratively calls the GPU kernels.

\begin{enumerate}[1.]

\item (Integrate): Update time-depend particle attributes. The position, velocity and angular velocity are updated.

\item (CalcHash): Assign each particle to the associated cell. The particle data has the attribute of the cell ID.

\item (BitonicSort): Sort the particle ID data in the cell ID order.

\item (FindCellBoundsAndReorder): Reorder all the data of the particle attributes in the cell ID order by using sorted particle IDs created in the "BitonicSort" kernel.

\item (ForceGravity): Add the gravitational force, $ m \boldsymbol{g}$, with $\boldsymbol{g}$ being the gravitational acceleration.

\item (InitializeContactIDs): Initialize the pair-wise variable, $\boldsymbol\delta_{t}$, which is a sparse matrix. If the two particles associated with an element of the matrix $\boldsymbol\delta_{t}$ are not in contact at the last contact check, this kernel delete the element.

\item (Collide): Find contact particles and calculate the DEM forces acting on the thread-associated particle.

\item (CollideRectangle): Check the contact between particles and rectangles of surfaces. If a particle and a surface are in contact, calculate the force acting between them.

\item (CollideLine): Check the contact between particles and lines. If a particle and a line are in contact, calculate the force acting between them.

\end{enumerate}

\subsection{Optimization of DEM force calculations}
The successive threads in a warp will get different results of the contact checks. For example, if one thread associated with a particle finds a contact particle and the others do not find contact particles, the others will wait for the one thread to finish the force calculations.
We cannot avoid the warp-divergence essentially, because we need the contact detections in DEM. Our procedure reduces the time which inactive CUDA cores cost to wait for the active thread to finish the calculations.

In the DEM, the probability that two particles selected from the neighbor cells are in contact is small. So, when the calculation of the contact forces is going on, the number of active threads in a warp is small.
In addition, the time to calculate the forces acting on particles is rather large in DEM. Since the majority of the threads in a warp wait for the minority to finish the calculations for the long time, the impact of warp-divergence becomes larger.

\begin{figure}[H]
\centering
\includegraphics[width=120mm]{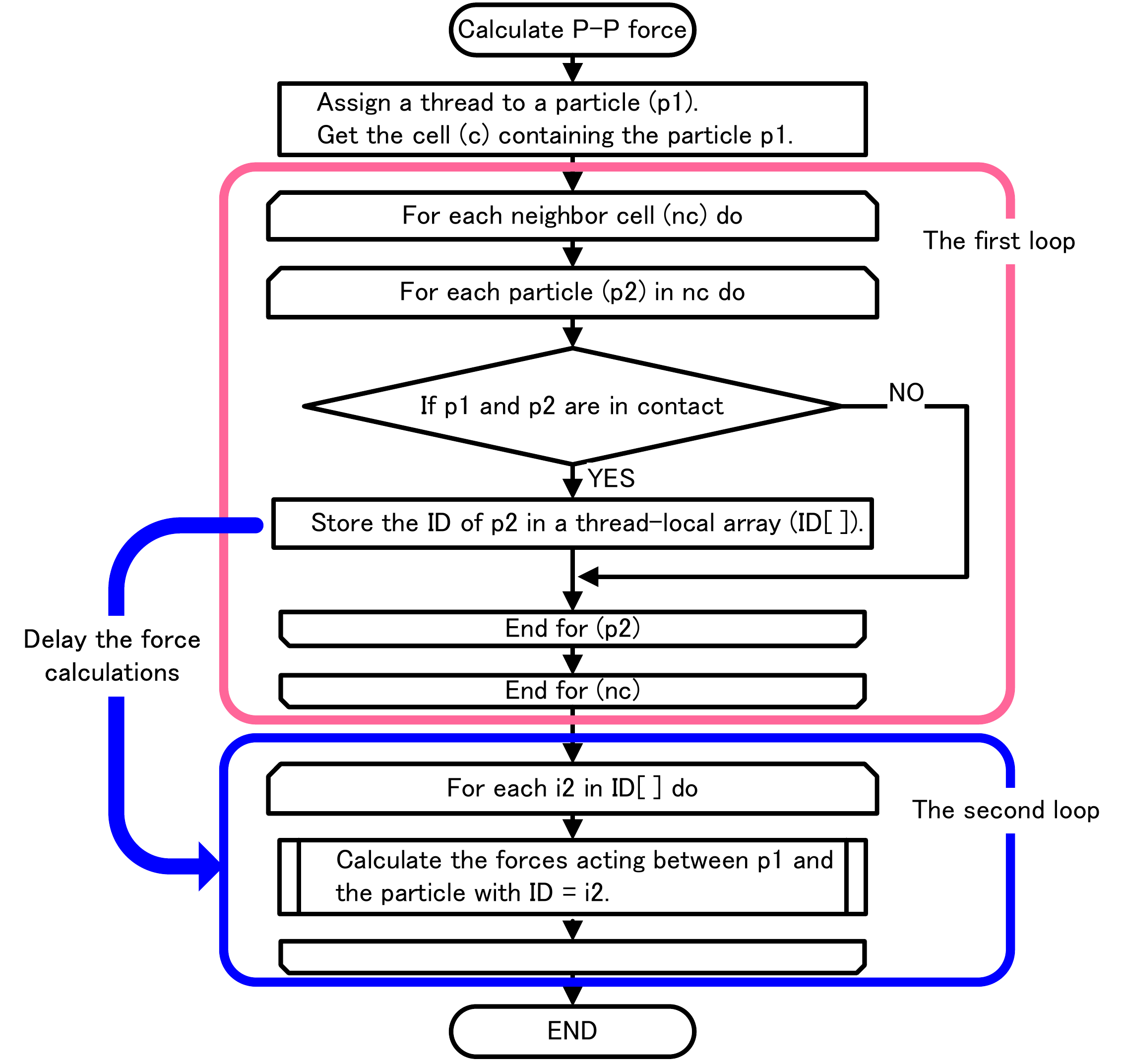}
\caption{A detailed flow chart, which shows how one thread calculates the forces acting on the thread-associated particle.}\label{figure-flow-chart}
\end{figure}

Fig.~\ref{figure-flow-chart} shows the detailed flow chart of our improved DEM force calculations. We modified the source code not to calculate the DEM force soon after the contact check. We moved the DEM force calculations into the second for-loop.

\section{Results}
\label{}

\subsection{Simulation platforms}
\label{}

The GPU we used in this simulation is M2050, whose architecture is called Fermi. This has 448 CUDA cores. The OpenCL library used to generate GPU executable binary file is included in the CUDA 4.1 developer's kit. The OpenCL source code is compiled with the level 3 optimization. The CPU in the host system is Intel X5670. This system has the OS, CentOS release 5.4.

\subsection{Performance evaluation}
\label{}

In order to test the performance, we initialized the particles in the way that particles are created on a uniform grid. We added small random numbers to the center positions of the created particles so that the distances of pairs of the particles are randomly distributed. The initial particles are not in contact. After initial particles were created, we executed the warm-up simulation until the particles become dense. The screenshots of Fig.~\ref{figure-simulation} show the configuration used to measure the performance.

\begin{figure}[H]
\centering
\includegraphics[width=150mm]{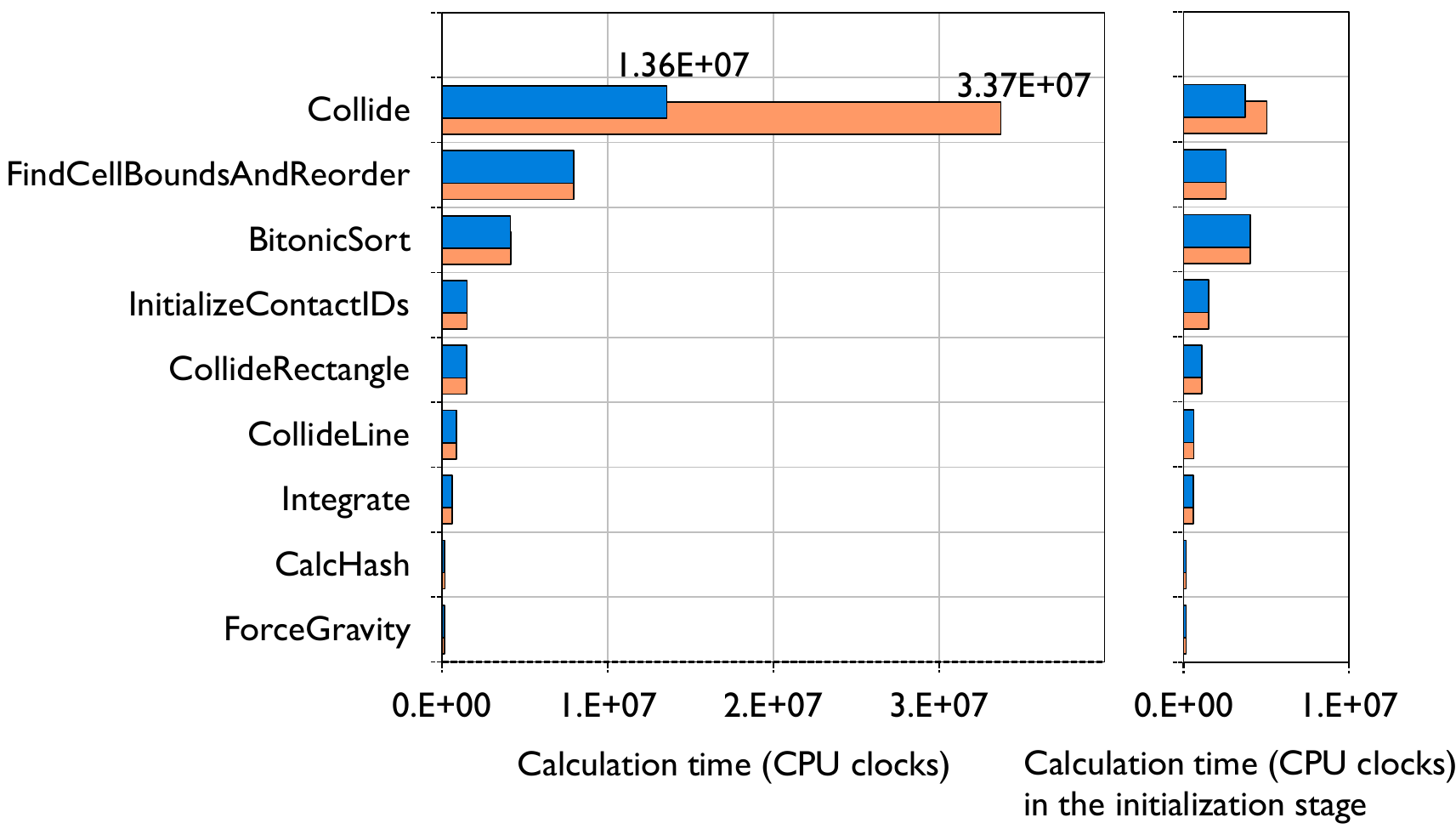}
\caption{Calculation time each kernel costs in the simulation. The left graph shows the calculation time measured after the particles become dense. The calculation time is measured in units of the CPU clocks on the host machine. The right graph shows the calculation time at the initialization stage, where particles are not in contact. The blue bars show the calculation times by the proposed method. The red bars show the calculation times by the previous one.}\label{figure-performance}
\end{figure}

As we have shown in Fig.~\ref{figure-performance}, the proposed method makes the performance of the calculations of the forces acting between particles, "Collide", 2.5 times faster. The number of particles used to measure this performance is 131072. The performance was measured in units of the CPU clocks. We checked that the speed-up ratio was comparable for all the numbers of particles greater than 0.1 million. We tested up to 2 million particles.

Among all the GPU kernels, the most time-consuming part is the calculations of the forces acting between particles, "Collide".
After reducing the time to execute "Collide", the time is comparable to that of the sorting, "BitonicSort" and "FindCellBoundsAndReorder".

\section{Conclusions}
\label{}
We focused on how the DEM force calculations are carried out on the GPU. We examined how each of CUDA cores works when the threads in a warp are executing the force calculations in the previous un-optimized code. We found that most CUDA cores are inefficiently used due to the warp divergences.

This paper suggested the strategy to reduce the impact of the warp divergences on the calculation time. According to the strategy, we modified the implementation so that the for-loop statement for the force calculations is divided into two for-loops. In the first for-loop, the program searches the particles actually in contact with the thread-associated particle and stores the IDs of the contact particles. In the second loop, the program concentrates on the DEM force calculations by using the searched particle IDs. We showed that the CUDA cores are more efficiently used in the second loop.

This modified code has an additional overhead. We have to store the IDs of the contact particles in the first loop and load them in the second loop. In spite of this, the time to calculate the forces is reduced to about $40 \%$ of the original time.

The strategy we have shown here is not very intrinsic to the DEM. This can be applied to similar kind of calculations which have checking the distance between two positions before calculating forces. But the speed-up depends on the kind of force calculations. This strategy becomes more effective if the force calculations cost more elapsed time or has larger amount of arithmetic operations.

\bibliographystyle{elsarticle-num}

\section*{\refname}
\label{}
\bibliography{dem-opencl-bib}

\end{document}